\documentclass[letterpaper]{article}
\usepackage[totalwidth=480pt, totalheight=680pt]{geometry}

\usepackage[T1]{fontenc}
\usepackage{amsfonts,amsmath,amssymb,amsthm}
\usepackage{algorithm}
\usepackage{algorithmicx}
\usepackage{algpseudocode}
\usepackage{booktabs}
\usepackage{tabularx}
\usepackage{multirow}
\usepackage{tikz}
\usepackage{url}
\usepackage{xurl}
\usepackage[round]{natbib}
\usepackage{xcolor}
\definecolor{darkgreen}{rgb}{0,0.5,0}
\usepackage{hyperref}
\hypersetup{
    colorlinks=true,
    linkcolor=red,
    citecolor=darkgreen,
    filecolor=magenta,
    urlcolor=blue
}
\usepackage{authblk}

\newtheorem{theorem}{Theorem}[section]

\newtheorem{lemma}[theorem]{Lemma}
\newtheorem{corollary}[theorem]{Corollary}
\theoremstyle{definition}
\newtheorem{definition}[theorem]{Definition}

\theoremstyle{remark}

\makeatletter
\providecommand*{\theHALG@line}{}
\renewcommand*{\theHALG@line}{\thealgorithm.\arabic{ALG@line}}
\makeatother

\newcommand{\eps}{\varepsilon}
\newcommand{\MMS}{\mathsf{MMS}}
\newcolumntype{Y}{>{\raggedright\arraybackslash}X}

\title{\bf Online Fair Division of Chores with Upgrade Recourse}
\author[1]{Yichen Huang}
\author[2]{Nicholas Teh}
\affil[1]{Harvard University, USA}
\affil[2]{University of Oxford, UK}
\date{}

\begin{document}

\maketitle

\begin{abstract}
We study the online fair division of indivisible chores among $n$ agents under \emph{upgrade recourse}, where earlier assignments may be revised but each chore previously assigned to an agent must be replaced by a distinct chore that she finds no more costly. We show that a simple adaptation of round robin maintains envy-freeness up to one chore (EF1) together with a $(2-1/n)$-approximation of the maximin share (MMS), and that no deterministic algorithm without predictions can do better, even when all agents rank the chores in the same order. This stands in contrast to the allocation of goods, where the optimal MMS factor vanishes as $n$ grows and cannot be achieved together with EF1. Moreover, knowing only the number of chores modulo a period of length $O(n)$ (and nothing about the costs) suffices to improve upon this factor at the final configuration, while the same algorithm retains a factor below $3$ even if this information is wrong. On the other hand, envy-freeness up to any chore (EFX) essentially requires the exact number of chores: uncertainty between two consecutive values already rules out every finite approximation, and even exact knowledge does not allow EFX to be combined with an MMS guarantee better than the trivial factor of $n$.
\end{abstract}

\section{Introduction}
According to GitHub's 2025 Octoverse report, developers on the platform merged an average of 43.2 million pull requests each month~\citep{github2025octoverse}---roughly a thousand every minute. Before it is merged, a pull request can be sent to teammates for review, and reviewing someone else's code is a textbook example of a \emph{chore}: a task that brings the person who performs it a cost rather than a benefit. GitHub can assign incoming review requests automatically, using either a round-robin rule, which alternates between the members of a team, or a load-balance rule, which favors members with fewer recent and outstanding requests~\citep{github2026review}. Customer service platforms such as Zendesk route support tickets by analogous rules, based on round robin or on each staff member's spare capacity~\citep{zendesk2026roundrobin}, and online assignment of heterogeneous tasks to workers has been studied for crowdsourcing markets such as Amazon Mechanical Turk~\citep{ho2012online}. In each of these settings, a web platform must decide, one request at a time, who does the work.

Two features of these applications make fair allocation difficult. First, work arrives \emph{online}: a request needs an owner as soon as it appears, long before the day's workload is known. Second, the same task can cost different people very different amounts of effort---a pull request that takes the author of the affected code ten minutes may take a newcomer an afternoon. The routing rules above are simple and transparent (indeed, our main algorithm builds on round robin), but they count requests rather than effort: GitHub's load-balance rule, for instance, aims for each team member to review ``an equal number of pull requests'' in any 30-day period~\citep{github2026review}. Moreover, since work arrives online, the fairest assignment of the morning's requests may depend on what arrives in the afternoon.

Fortunately, assignments on these platforms are not final: a review can be passed to another reviewer, and a ticket can be handed to a colleague. Reassignment is a mixed blessing, however. Unrestricted reassignment can burden a person with work she finds far more costly than what she originally accepted---a reviewer who agreed to check a two-line fix will not welcome finding it swapped for a sprawling refactor (readers who have been handed a last-minute conference review will know the feeling). Prohibiting reassignment altogether avoids this problem but, as we discuss next, severely limits fairness. Our key question is this: \emph{How fairly can chores be allocated online if assignments may be revised, but only in ways that do not make any agent's earlier commitments more costly for her?}

The online fair division model requires each arriving item to be allocated immediately~\citep{AleksandrovEtAl2015,AleksandrovWalsh2020Survey}. For $n$ agents with additive costs, an agent's maximin share (MMS)~\citep{budish2011combinatorial,aziz2017chores} is the smallest cost she can guarantee herself by dividing the chores into $n$ bundles and receiving the most costly one, evaluated using her own costs. An allocation is $\alpha$-MMS if it gives every agent cost at most $\alpha$ times her maximin share; smaller factors are better. With irrevocable assignments and an unknown horizon (i.e., the number of chores), the trivial $n$-MMS guarantee is optimal up to an arbitrarily small additive term~\citep{song2025onlinechores}. In other words, if assignments are final, no online algorithm can guarantee substantially better than handing every chore to a single (unlucky) agent. Dynamic fair division permits changes and typically measures the number of disruptions~\citep{friedman2015dynamic,HeProcacciaPsomasZeng2019}. Our concern is different: which changes are acceptable to the affected agents?

\citet{huang2026upgrade} recently proposed \emph{upgrade recourse} for goods, requiring every reassignment to leave each affected agent at least as well off. In this paper, we initiate its study for chores. Note that upgrade recourse for chores must be defined carefully and differently from that for goods, since it is not possible to leave every agent at least as well off when a chore arrives. Whenever a chore arrives, every previously assigned chore must be matched to a distinct chore in its owner's new bundle that she finds no more costly. Thus each earlier commitment has a separate replacement, and exactly one agent receives an additional chore, not necessarily the arriving one. In the code review example, the platform may swap a pending review for one that the reviewer finds no more costly, but never for a more demanding one, nor may it replace two of her pending reviews with a single review that costs her more than either. We place no bound on the number of reassignments, and we do not model when chores are performed, so every assigned chore remains available for reassignment (think of reviews that nobody has started yet).

Our algorithms use the \emph{insertion} procedure of \citet{huang2026upgrade}, adapted to chores. Fix a sequence of picking turns, each owned by an agent. When a chore arrives, the owner of each old turn, in order, keeps the less costly of the chore in that turn and the chore passed to her (initially the arrival), and passes the other on; the chore passed on from the last old turn is placed in a new final turn. Every old turn thus weakly improves for its owner. More importantly, the result is exactly the \emph{least-cost picking} allocation, in which the owners of the turns, in order, each take a least costly remaining chore among all chores observed so far. The problem therefore separates into implementing a fixed sequence online and choosing one with suitable fairness guarantees. With a round-robin sequence, insertion gives each agent exactly as many chores as round-robin routing would, but lets the agents' costs decide who holds which chore. Notably, it never needs cost magnitudes: each step asks an agent only which of two chores she finds less costly.

This separation also explains the value of a small prediction. For goods, the $k$th picking turn guarantees its owner her $k$th most valuable good or better, whatever the horizon~\citep{huang2026upgrade}. For chores, the guarantee of a turn depends instead on how many turns follow it---the last turn receives the one remaining chore, which may be the most costly one. Good picking sequences for chores therefore assign each \emph{cost rank} (rank~$1$ for the most costly chore, rank~$2$ for the next, and so on) to an agent and let the agents pick in reverse order of rank~\citep{aziz2024ordinalchores}. If this assignment follows a periodic pattern, knowing the final number of chores modulo the period tells the algorithm where in the pattern to start. Thus $O(\log n)$ bits of horizon information suffice to surpass what any prediction-free algorithm achieves. Some platforms already forecast their workload; Zendesk, for instance, forecasts inbound ticket volume up to a year ahead from historical data~\citep{zendesk2026wfm}. We ask for far less (only the number of chores modulo a period of length $O(n)$) but we ask for it exactly. For this reason, we also bound the loss when the prediction is wrong; in no case do we need to predict future costs.

\subsection{Our Results}
Costs are additive throughout, and algorithms are deterministic unless stated otherwise. Table~\ref{tab:results} summarizes the MMS guarantees.

In Section~\ref{sec:insertion}, we observe that insertion~\citep{huang2026upgrade} has upgrade recourse and maintains the least-cost picking allocation of any fixed picking sequence (Lemma~\ref{lem:serial-insertion}). Insertion with a round-robin sequence, which we call \emph{round-robin insertion}, maintains \emph{envy-freeness up to one chore} (EF1)~\citep{budish2011combinatorial,aziz2022goodschores} (any envy that an agent has toward another agent can be eliminated by removing some chore from her own bundle) together with $(2-1/n)$-MMS (Theorem~\ref{thm:item-chore-ef1}).

Section~\ref{sec:lower-bounds} shows that the factor $2-1/n$ is optimal for every $n\ge2$ (Theorem~\ref{thm:chore-mms-lower-bound}), even when all agents have the same strict ranking of the chores. The bound holds without requiring EF1 or restricting algorithms to be \emph{ordinal}, i.e., to use only rankings rather than numerical costs. In other words, without horizon information, no algorithm with upgrade recourse improves in the worst case on this simple modification of round-robin routing. We also give lower bounds under \emph{bundle-level upgrades}, which protect total cost rather than each earlier chore separately: round-robin insertion remains optimal for two and three agents, while for $n\ge4$ the optimal factor lies between $\sqrt3$ and $2-1/n$.

In Section~\ref{sec:predictions}, we show that knowing the residue of the number of chores modulo $L=2n-\lfloor n/2\rfloor$ yields an MMS factor of $1+(n-1)/L$, strictly improving on the optimal prediction-free factor. The same algorithm retains a factor below $3$ under arbitrary predictions (Theorem~\ref{thm:chore-residue-prediction}). More generally, we determine the optimal tradeoff between the factor under correct predictions, called \emph{consistency}, and that under arbitrary predictions, called \emph{robustness}, among fixed patterns of length $L$ (Theorem~\ref{thm:periodic-tradeoff}). The tradeoff depends on how far apart the pattern places the two turns of each agent who appears twice. In this class, however, the robustness is always at least $2$, so when the prediction is wrong, the guarantee is worse than that of round-robin insertion, which ignores the prediction.

Section~\ref{sec:efx} gives a contrasting characterization for \emph{envy-freeness up to any chore} (EFX)~\citep{caragiannis2019unreasonable}, which requires any envy that an agent has toward another agent to disappear as soon as any chore is removed from her own bundle. When the horizon $m$ is known exactly, the optimal EFX factor is $\max\{1,m-n\}$, but uncertainty between $m=n$ and $m=n+1$ already makes every finite factor impossible. We also determine how many distinct messages about the horizon are needed for a finite EFX factor, and show that even exact knowledge of the horizon cannot reconcile a finite EFX factor with any MMS factor below $n$ that does not depend on $m$. These EFX lower bounds remain valid under bundle-level upgrades. In other words, a platform that wants this stronger guarantee must essentially know how much work will arrive, and even then some instances force it to give the first $m-n+1$ chores to a single team member.

\paragraph{Comparison with goods.}
Insertion and both recourse rules of \citet{huang2026upgrade}, which protect either each earlier item (\emph{item-level} upgrades) or only each bundle (bundle-level upgrades), carry over to chores, the bundle-level rule with a modification (see Section~\ref{sec:model}). However, the answers differ markedly from those for goods. For goods under item-level upgrades, the optimal MMS factor is essentially $1/H_n$, where $H_n$ is the $n$th harmonic number, so it vanishes as $n$ grows, and requiring EF1 lowers it to $1/n$; for chores, the optimal factor stays below $2$ and is attained together with EF1. For goods, bundle-level upgrades turn the vanishing factor into a constant and permit exact MMS for two agents, as well as exact EFX for two agents and for identical valuations; for chores, they do not improve the MMS factor for two or three agents, and they do not make any finite EFX factor possible without horizon information. Finally, for goods, predictions of every agent's ranked values transfer offline MMS guarantees online, whereas for chores, $O(\log n)$ bits about the horizon, and none about costs, beat the optimal prediction-free factor.

\begin{table}[t]
\caption{MMS factors for deterministic algorithms; smaller is better. For comparison, the optimal factor with irrevocable assignments is essentially $n$~\citep{song2025onlinechores}. Bold factors are optimal. The two residue rows use the same algorithm; among fixed patterns of length $L=2n-\lfloor n/2\rfloor$, its consistency is optimal, and so is its robustness given that consistency (Theorem~\ref{thm:periodic-tradeoff}), but we do not claim optimality over all algorithms.}
\label{tab:results}
\begin{tabularx}{\textwidth}{@{}l l c c l Y@{}}
\toprule
Recourse & Information & Agents & MMS factor & Holds at & Result \\
\midrule
Item-level & None & $n\ge2$ & $\boldsymbol{2-1/n}$ & Every prefix & Round-robin insertion; Theorems~\ref{thm:item-chore-ef1} and~\ref{thm:chore-mms-lower-bound}. \\
Item-level & Correct horizon residue & $n\ge2$ & $1+(n-1)/L$ & Final horizon & $4/3$ for $n=2$, $7/5$ for $n=3$, below $5/3$ otherwise; Theorem~\ref{thm:chore-residue-prediction}. \\
Item-level & Arbitrary horizon residue & $n\ge2$ & $<3$ & Every prefix & Exactly $2+2(n-\lfloor n/4\rfloor-2)/L$; Theorem~\ref{thm:chore-residue-prediction}. \\
Bundle-level & None & $2,3$ & $\boldsymbol{2-1/n}$ & Every prefix & Round-robin insertion; Theorems~\ref{thm:item-chore-ef1} and~\ref{thm:bundle-mms-lower-bound}. \\
Bundle-level & None & $n\ge4$ & $[\sqrt3,2-1/n]$ & Every prefix & Optimal factor open; Theorem~\ref{thm:bundle-mms-lower-bound}. \\
\bottomrule
\end{tabularx}
\end{table}

\subsection{Related Work}
We discuss the most closely related work here; Appendix~\ref{app:further-related-work} gives a more detailed discussion, including recourse in online algorithms, offline chore allocation, ordinal scheduling, and temporal voting.

\citet{AleksandrovEtAl2015} studied mechanisms for online fair division of indivisible items, motivated by food banks; \citet{AleksandrovWalsh2020Survey} survey the area. \citet{benade2018envy} show that irrevocable assignments of goods can keep envy within $\tilde O(\sqrt{T/n})$ after $T$ items, which is asymptotically optimal, and \citet{zeng2020fairness} study the tradeoff between fairness and efficiency in this setting. \citet*{zhou2023multi} study online MMS allocation with additional information, while \citet{song2025onlinechores} establish a tight lower bound for chores with irrevocable assignments. For goods, randomization against an oblivious adversary does not ensure MMS guarantees with high probability: every randomized algorithm has an input on which it returns an $\alpha$-MMS allocation with probability at most $5/6$, for every $\alpha>0$~\citep{cohen2026oblivious}. In temporal fair division, \citet{elkind2025temporal} require EF1 at every prefix with irrevocable assignments, but assume that all items are known in advance, and \citet{choi2026temporal} identify tractable cases for mixed manna, in which an item may be a good for some agents and a chore for others; under upgrade recourse, round-robin insertion maintains EF1 at every prefix without this knowledge. \citet{amanatidis2026Online} study online fair division with a reordering buffer. Online recourse has also been studied in matching~\citep{bernstein2019online}, scheduling and flows~\citep{gupta2014maintaining}, and makespan scheduling with bounded migration~\citep{sanders2009online}, and it is inherent in online house allocation, where envy-freeability with subsidies can be maintained using bounded recourse~\citep{teh2026house}. These works bound the amount of reassignment; upgrade recourse restricts its direction from each agent's perspective. The idea of non-worsening recourse has been considered in online metric embeddings~\citep{Coester2026Online}. Picking sequence monotonicity has been studied in weighted fair division~\citep{chakraborty2021picking}, and insertion uses it to implement picking sequences online~\citep{huang2026upgrade}; weighted picking sequences for chores have also been used to allocate weighted mixed manna~\citep{teh2026weightedmixed}.

Our results with predictions build on the ordinal results for chores of \citet*{aziz2024ordinalchores} and on the study of picking sequences by \citet{feigehuang2023chores}. We do not claim new offline approximation factors; our contribution is the online implementation with limited horizon information, the strict improvement over prediction-free algorithms, and the optimal tradeoff for fixed patterns of length $L$. Algorithms with predictions have been studied for caching~\citep{lykouris2021competitive} and online allocation~\citep{banerjee2022online,cohen2023general}; our prediction concerns only the starting position of a periodic sequence. Correct predictions can also be viewed as $O(\log n)$ bits of online advice~\citep{boyar2017online}. With irrevocable assignments of goods, \citet*{neoh2026additional} lift offline share-based guarantees online using predictions of each agent's multiset of values; with the same predictions, \citet{neoh2026frequency} show that a positive approximation of EF1 is possible if and only if $n\le4$, and rule out any positive approximation of EFX for three or more agents. \citet{choo2026proportionality} and \citet{neoh2026closing} approximate proportionality up to one good using the maximum item value, and \citet{melissourgos2025onlineefx} study online EFX in the same setting with predictions of valuation vectors. Predictions have also been used for budget-constrained online fair division and online house allocation~\citep{cohen2026budget,teh2026house}. Our EFX results instead quantify the horizon information needed under upgrade recourse.

For identical costs, MMS equals the optimal makespan on identical machines, for which Graham's list scheduling~\citep{graham1966bounds} is $(2-1/n)$-competitive. If only the order of the job sizes is known, \citet*{liu1996ordinal} give an ordinal algorithm with factor $1+(m-1)/(m+\lceil m/2\rceil)$ on $m$ machines, which is optimal for $m=2,3$; for identical costs, our consistency $1+(n-1)/L$ with $L=n+\lceil n/2\rceil$ equals this factor. Our lower bounds for bundle-level upgrades use work of \citet*{faigle1989partition}, \citet{rudin2003online}, and \citet*{braun2025four}. We verify their validity under this weaker rule and extend the recursive four-machine construction to every $n\ge4$. \citet*{sun2023fairness} study offline relationships among chore-fairness notions; in particular, EF1 implies $(2-1/n)$-MMS.

\section{Model and Upgrade Recourse}
\label{sec:model}
There are $n\ge2$ agents, denoted $[n]=\{1,\dots,n\}$, and chores $D=\{d_1,\dots,d_m\}$ arriving in that order. Let $D^{(t)}=\{d_1,\dots,d_t\}$ and $D^{(0)}=\varnothing$. Agent $i$ has a nonnegative cost function $c_i:2^D\to\mathbb R_{\ge0}$ and prefers lower cost. Costs are additive, i.e., $c_i(S)=\sum_{d\in S}c_i(d)$, where we write $c_i(d)$ instead of $c_i(\{d\})$ for a single chore~$d$. All agents' costs for a chore are revealed when it arrives.

At every time $t$, the algorithm immediately outputs a complete allocation $\mathcal A^{(t)}=(A_1^{(t)},\dots,A_n^{(t)})$ partitioning $D^{(t)}$. Unless a prediction is explicitly supplied, the horizon $m$ and future costs are unknown. An adaptive adversary may choose subsequent arrivals after seeing earlier allocations. For deterministic algorithms this is equivalent to requiring the guarantee on every fixed finite sequence. We write $\mathcal A=\mathcal A^{(m)}$.

Fairness is evaluated for the final allocation~$\mathcal A$; without predictions, a guarantee for every finite sequence must hold at every prefix, since arrivals could stop there. By contrast, recourse must hold at \emph{every} update, even when predictions are wrong.

\subsection{Fairness}
An allocation is \emph{envy-free} if $c_i(A_i)\le c_i(A_j)$ for every $i,j$. For $\alpha\ge1$, it is \emph{$\alpha$-EF1} if, for every $i,j$ with $A_i\ne\varnothing$, some $d\in A_i$ satisfies $c_i(A_i\setminus\{d\})\le\alpha c_i(A_j)$.
It is \emph{$\alpha$-EFX} if this inequality holds for every $d\in A_i$, including zero-cost chores. We omit $\alpha$ when it equals~$1$.

Let $\Pi_n(S)$ be the set of partitions of $S$ into $n$ possibly empty bundles. Then, agent $i$'s maximin share is defined as $\MMS_i(S)=\min_{(P_1,\dots,P_n)\in\Pi_n(S)} \max_{j\in[n]}c_i(P_j)$.
An allocation of $S$ is \emph{$\alpha$-MMS} if $c_i(A_i)\le\alpha\MMS_i(S)$ for all $i$. We will use
\begin{equation}
\label{eq:mms-basic}
 c_i(S)\le n\MMS_i(S),\quad
 c_i(d)\le\MMS_i(S)\quad \text{for } d\in S.
\end{equation}
Both inequalities follow from any MMS partition.

\subsection{Upgrade Recourse}
\begin{definition}[Upgrade recourse]
\label{def:item-upgrades}
An update at time $t$ is \emph{upgrade-feasible} if, for every agent $i$, there exists an injection $\phi_i:A_i^{(t-1)}\to A_i^{(t)}$ such that $c_i(\phi_i(d))\le c_i(d)$ for every $d\in A_i^{(t-1)}$. An algorithm has \emph{upgrade recourse} if every update is upgrade-feasible.
\end{definition}
This is the chore counterpart of item-level upgrade recourse for goods~\citep{huang2026upgrade}. In other words, each earlier chore of an agent has a separate replacement that she finds no more costly, possibly the chore itself. Since injections compose, the same holds between the allocations at any two times. In particular, each agent's number of chores is nondecreasing, and in each update exactly one agent receives an additional chore, though not necessarily the arriving one. For example, suppose that an old chore of common cost $5$ belongs to agent $1$ and a chore of cost $1$ arrives. Giving the arrival to agent $1$ and the old chore to agent $2$, whose bundle was previously empty, is upgrade-feasible, since agent $1$'s old chore can be matched to the arrival.

Summing the inequalities in Definition~\ref{def:item-upgrades} yields the following consequence, which is all that most of our lower bounds require.

\begin{lemma}
\label{lem:upgrade-sums}
In every upgrade-feasible update at time $t$, there exist an agent $r$ and a chore $h\in A_r^{(t)}$ such that $c_i(A_i^{(t)})\le c_i(A_i^{(t-1)})$ for all $i \neq r$, and $c_r(A_r^{(t)}\setminus\{h\})\le c_r(A_r^{(t-1)})$.
\end{lemma}
\begin{proof}
Since the bundles $A_i^{(t)}$ are disjoint and each $\phi_i$ is injective, the images of the injections consist of $t-1$ distinct chores, so exactly one chore $h\in D^{(t)}$ lies outside them; let $r$ be its owner. For $i\ne r$, the image of $\phi_i$ is all of $A_i^{(t)}$, so summing $c_i(\phi_i(d))\le c_i(d)$ over $d\in A_i^{(t-1)}$ gives the first inequality. For $r$, the image of $\phi_r$ is $A_r^{(t)}\setminus\{h\}$, and the same summation gives the second.
\end{proof}
In the example above, $r=2$ and $h$ is the old chore rather than the arrival.

\paragraph{Bundle-level upgrades.}
\label{sec:bundle}
A weaker rule requires only the two inequalities of Lemma~\ref{lem:upgrade-sums} in each update. We call this \emph{bundle-level upgrade recourse}, adapting the rule of that name for goods, which requires that no agent's value decrease~\citep{huang2026upgrade}. The literal chore analogue, that no agent's cost increase, cannot be guaranteed: the first chore raises its recipient's cost whenever every agent finds it costly. Our rule therefore lets one agent's cost rise by at most her cost for one chore in her new bundle, while no other agent's cost increases. It protects total cost rather than each earlier chore separately, and is hence weaker. We give MMS lower bounds for the weaker rule in Section~\ref{sec:bundle-mms}; the EFX results in Section~\ref{sec:efx} also hold under it.

\section{Maintaining Fairness by Insertion}
\label{sec:insertion}
Insertion was introduced for goods by \citet{huang2026upgrade}. We restate it for least-cost picking and include the short proof for completeness. Fix an infinite sequence $\sigma\in[n]^{\mathbb N_+}$. Its least-cost picking allocation on $t$ chores lets agents $\sigma_1,\dots,\sigma_t$ successively choose a least costly remaining chore from the \emph{entire} observed set. The chore picked at turn $k$ is held in \emph{slot}~$k$ and need not be the one arriving at time~$k$. Ties are broken by a fixed deterministic order whose restriction to previously observed chores never changes. Write $b\succ_i o$ when agent $i$ prefers $b$ to $o$ under this rule, which implies $c_i(b)\le c_i(o)$.

\begin{algorithm}[H]
\caption{Chore Insertion}
\label{alg:serial-insertion}
\begin{algorithmic}[1]
\Require An infinite picking sequence $\sigma$.
\State Initially there are no slots.
\For{each arriving chore $d_t$}
    \State Reveal $(c_i(d_t))_{i\in[n]}$; set buffer $b\gets d_t$.
    \For{$k=1,\dots,t-1$}
        \State Let $i\gets\sigma_k$ and let $o$ occupy slot $k$.
        \If{$b\succ_i o$}
            \State Put $b$ in slot $k$; set $b\gets o$.
        \EndIf
    \EndFor
    \State Create slot $t$ with owner $\sigma_t$ and chore $b$.
    \State Give each agent the chores in her slots.
\EndFor
\end{algorithmic}
\end{algorithm}

\begin{lemma}[Insertion invariant]
\label{lem:serial-insertion}
For every fixed $\sigma$, Algorithm~\ref{alg:serial-insertion} has upgrade recourse. At every time $t$, it maintains exactly the least-cost picking allocation induced by the first $t$ turns on $D^{(t)}$.
\end{lemma}
\begin{proof}
Every old slot keeps its incumbent or receives a weakly less costly chore for its owner. Since distinct slots hold distinct chores, matching each old chore to the new occupant of its slot gives the required injections.

For the picking claim, we proceed by induction on $t$; the case $t=1$ is immediate. Consider time $t\ge2$, and suppose that slots $1,\dots,k-1$ already hold the first $k-1$ picks on $D^{(t)}$. Then the chores remaining before turn $k$ are the buffer together with the old occupants of slots $k,\dots,t-1$, and by the induction hypothesis, the latter are exactly the chores remaining before turn $k$ in the picking allocation on $D^{(t-1)}$. Since the tie-breaking order among old chores is unchanged, the old occupant of slot $k$ is still its owner's preferred chore among those old chores. Her preferred chore among all remaining chores is therefore either that occupant or the buffer. The algorithm keeps precisely this choice in slot $k$ and passes the other onward. Applying this argument for $k=1,\dots,t-1$ in turn, we see that the last buffer is the only remaining chore, which turn $t$ receives in slot $t$.
\end{proof}
If observed costs are stored, update $t$ uses $t-1$ comparisons and $O(n+t)$ time, for a total of $O(nm+m^2)$. As many as $t-1$ old chores may change owners: for example, if the agents have identical costs that strictly decrease with arrival order and $\sigma$ is round-robin, every old chore moves to the next slot and hence to another agent. Upgrade recourse restricts the nature, not the amount, of reassignment.

We now apply insertion to a \emph{round-robin} sequence, which repeats a fixed permutation of $[n]$.

\begin{theorem}
\label{thm:item-chore-ef1}
Chore insertion with any round-robin sequence has upgrade recourse and maintains EF1 and $(2-1/n)$-MMS.
\end{theorem}
\begin{proof}
Recourse follows from Lemma~\ref{lem:serial-insertion}. By the same lemma, the allocation at each time is the least-cost picking allocation of the round-robin sequence on the chores observed so far, so it suffices to consider this allocation on~$D$. Fix agents $i\ne j$, and list their selected chores in the order of their turns as $x_1,\dots,x_a$ and $y_1,\dots,y_b$. If $a=0$ there is nothing to prove; otherwise, we show that removing $x_a$ suffices. When $i$ precedes $j$ in the repeated order, $b\ge a-1$ and $y_\ell$ is available when $i$ selects $x_\ell$ for $\ell\le a-1$. Hence $c_i(A_i\setminus\{x_a\})\le\sum_{\ell=1}^{a-1}c_i(y_\ell)\le c_i(A_j)$. When $j$ precedes $i$, $b\ge a$ and $y_{\ell+1}$ is available at $i$'s $\ell$th turn, giving the same inequality with the sum over $y_2,\dots,y_a$. Thus the allocation is EF1.

For completeness, we derive the standard MMS implication~\citep{sun2023fairness}. Since every chore costs agent~$i$ at most $\MMS_i(D)$ by~\eqref{eq:mms-basic}, EF1 gives $c_i(A_j)\ge c_i(A_i)-\MMS_i(D)$ for every $j\ne i$ (trivially if $A_i=\varnothing$). Summing over $j\ne i$ yields $c_i(D)\ge n\,c_i(A_i)-(n-1)\MMS_i(D)$, and $c_i(D)\le n\MMS_i(D)$ then gives $c_i(A_i)\le(2-1/n)\MMS_i(D)$.
\end{proof}
The EF1 argument is the classical round-robin argument~\citep{caragiannis2019unreasonable,aziz2022goodschores}, which also applies to goods under upgrade recourse~\citep{huang2026upgrade}. For goods, however, EF1 guarantees only a $1/n$ fraction of MMS, and under item-level upgrades no algorithm maintaining EF1 does better, so the near-optimal goods algorithms use unbalanced picking sequences~\citep{huang2026upgrade}. For chores, Section~\ref{sec:lower-bounds} shows that the factor $2-1/n$ implied by EF1 is already optimal.

\section{Limits Without Predictions}
\label{sec:lower-bounds}
We now show that round-robin insertion attains the best possible MMS factor, even when an algorithm uses numerical costs and need not maintain EF1.

\subsection{The Optimal MMS Factor}
\begin{theorem}
\label{thm:chore-mms-lower-bound}
For every $n\ge2$, the optimal MMS factor of deterministic prediction-free algorithms with upgrade recourse is $2-1/n$. The lower bound holds even when every agent's singleton costs are strictly positive and strictly increase with arrival order.
\end{theorem}
The construction treats agent~$1$ differently from the other agents. For each of agents $2,\dots,n$, receiving a chore makes another assignment to her within the next $n-1$ arrivals incompatible with an MMS factor below $2-1/n$. Agent~$1$ must therefore receive sufficiently many chores, each costing her almost~$1$. We then present at most $n$ more chores, one of which she must receive. Keeping more of the other agents available for these last chores requires assigning more of the earlier chores to agent~$1$, and the proof shows that, however the algorithm balances these two effects, her MMS ratio eventually exceeds every $\beta<2-1/n$.

\begin{proof}
The upper bound follows from Theorem~\ref{thm:item-chore-ef1}. For the lower bound, suppose for contradiction that an algorithm guarantees $\beta$-MMS for some $1\le\beta<2-1/n$.

First observe, as for goods whose values strictly decrease with arrival order~\citep{huang2026upgrade}, that old chores cannot change owners if every agent's costs strictly increase with arrival order. Indeed, since the images of the injections in Definition~\ref{def:item-upgrades} are disjoint, induction on $s$ shows that the image of each old chore $d_s$ is $d_s$ itself: all earlier chores are already used as images, and all later chores are more costly. We construct costs with this property, so assignments are irrevocable throughout the construction.

\paragraph{The initial chores.}
Give agent~$1$ cost $c_1(d_t)=1-2^{-t}$. For each agent $i\in\{2,\dots,n\}$, choose costs as follows. If she has not received a chore in the preceding $n-1$ arrivals, set $c_i(d_t)=4n(c_i(D^{(t-1)})+1)$. If she receives a chore $d_\tau$ whose cost is set by this rule, write $Q=c_i(d_\tau)$ and give the next $n-1$ chores costs $c_i(d_{\tau+r})=Q(1+r/(4n(n-1)))$ for $1\le r\le n-1$. After these arrivals, use the first rule again. We stop if she receives another chore during these $n-1$ arrivals. The costs increase strictly: the first rule exceeds the total cost of all preceding chores, and the second gives strictly increasing costs above~$Q$.

Such a second assignment would contradict the guarantee. To see this, consider the first time it occurs, say at $\tau+r$. All chores before $\tau$ have total cost less than $Q/(4n)$ to agent~$i$. The $r+1\le n$ chores from $\tau$ through $\tau+r$ each cost her at most $Q(1+1/(4n))$. Putting these chores in distinct bundles and adding all earlier chores to one of those bundles gives $\MMS_i(D^{(\tau+r)})<Q(1+1/(2n))$. Agent~$i$ herself has cost at least $2Q$, so her MMS ratio is greater than $2/(1+1/(2n))=4n/(2n+1)>2-1/n>\beta$. Consequently, each of agents $2,\dots,n$ receives at most one chore in any $n$ consecutive arrivals. In particular, every $n$ consecutive arrivals include an assignment to agent~$1$.

\paragraph{Choosing an initial prefix.}
Consider continuing these initial costs indefinitely. The infinite continuation only identifies this finite prefix: without a horizon prediction, the algorithm makes the same decisions on that prefix when we change the subsequent arrivals. Let $\ell_t$ be the number of consecutive assignments to agents $2,\dots,n$ at the end of the first $t$ arrivals. We have $0\le\ell_t\le n-1$. Let $\ell$ be the largest value that occurs infinitely often. Eventually $\ell_t\le\ell$, so every $\ell+1$ consecutive sufficiently late arrivals include an assignment to agent~$1$. Hence $\liminf_{t\to\infty}|A_1^{(t)}|/t\ge1/(\ell+1)$. There are arbitrarily large prefixes of length $M$ with $\ell_M=\ell$, and at each of them $c_1(A_1^{(M)})>|A_1^{(M)}|-1$, since $\sum_{t=1}^M2^{-t}<1$. Choose such an $M\ge(n-1)^2$; we will take $M$ sufficiently large below.

\paragraph{The last chores.}
Keep the cost rules for agents $2,\dots,n$ unchanged and append at most $n-\ell$ chores. We choose their costs for agent~$1$ so that each new chore alone determines her MMS when it arrives. Specifically, set
\begin{equation}
\label{eq:chore-last-costs}
 p_j:=c_1(d_{M+j})
 =\left\lceil\frac{M+\sum_{h<j}p_h}{n-1}\right\rceil
 \quad \text{for }  1\le j\le n-\ell.
\end{equation}
Since $M\ge(n-1)^2$, we have $p_1\ge n-1$. By induction, $p_{j+1}\ge p_j+1$ for every $j$, because the argument of the ceiling in~\eqref{eq:chore-last-costs} grows by $p_j/(n-1)\ge1$ from $j$ to $j+1$. Thus these costs strictly increase and exceed all initial costs, so no old chore can change owners. If none of these $n-\ell$ chores went to agent~$1$, they and the last $\ell$ initial chores would give $n$ consecutive assignments to the other $n-1$ agents, which we have ruled out. Stop when agent~$1$ first receives one of these chores, say $d_{M+j}$.

To compute agent~$1$'s MMS, put $d_{M+j}$ alone and put each of the $j-1\le n-1$ preceding appended chores in a different one of the remaining $n-1$ bundles. Since all $p_h$ are integers, each of these $n-1$ bundles can still accommodate a nonnegative integer amount of cost, namely $p_j$ minus its current cost, and by~\eqref{eq:chore-last-costs} these amounts sum to $(n-1)p_j-\sum_{h<j}p_h\ge M$. Because each initial chore costs agent~$1$ less than~$1$, a bundle that can accommodate an integer amount $a$ can receive any $a$ initial chores without exceeding $p_j$. Hence all $M$ initial chores fit into these bundles. The singleton cost $p_j$ gives the reverse inequality, so $\MMS_1(D^{(M+j)})=p_j$. Agent~$1$ retains her initial chores and has received no earlier appended chore. Her MMS ratio is therefore greater than $1+(|A_1^{(M)}|-1)/p_j\ge1+(|A_1^{(M)}|-1)/p_{n-\ell}$.

It remains to compare $p_{n-\ell}$ with the number of initial chores assigned to agent~$1$. Using $\lceil x\rceil\le x+1$ in~\eqref{eq:chore-last-costs} and induction gives $p_j\le(n/(n-1))^{j-1}(M/(n-1)+1)$; indeed, substituting these bounds for the preceding $p_h$ gives a geometric sum with the stated value. Combining this bound for $j=n-\ell$ with $\liminf_{t\to\infty}|A_1^{(t)}|/t\ge1/(\ell+1)$, we find that, as $M\to\infty$ along the chosen prefixes, the limit inferior of $1+(|A_1^{(M)}|-1)/p_{n-\ell}$ is at least
\[
 1+\frac{n-1}{\ell+1}
 \left(1-\frac1n\right)^{n-\ell-1}
 \ge 2-\frac1n,
\]
where the last inequality follows from Bernoulli's inequality $(1-1/n)^{n-\ell-1}\ge1-(n-\ell-1)/n=(\ell+1)/n$. Since $\beta<2-1/n$, a sufficiently large finite prefix and the appended chores give a ratio greater than $\beta$, a contradiction. 
\end{proof}

\subsection{Bundle-Level Upgrades}
\label{sec:bundle-mms}
The preceding proof uses a separate replacement for each earlier chore. Under bundle-level upgrades, an agent may exchange several chores for a single more costly chore without increasing her total cost, so the preceding argument does not apply. However, with identical costs that are nondecreasing in arrival order, the agents' costs still evolve as if no chore changed hands.

\begin{lemma}[Nondecreasing costs]
\label{lem:nondecreasing-chore-update}
Let $c$ be the agents' common cost function. If the chore $d_t$ has cost $p$ and every previous chore costs at most $p$, then every update at time $t$ satisfying the bundle-level rule has an agent $r$ such that
\begin{equation}
\label{eq:nondecreasing-chore-update}
 c(A_r^{(t)})=c(A_r^{(t-1)})+p\quad\text{and}\quad
 c(A_i^{(t)})=c(A_i^{(t-1)})\quad \text{for } i\ne r.
\end{equation}
\end{lemma}
\begin{proof}
Let $r$ and $h$ satisfy the two inequalities defining bundle-level upgrades. Then $c(A_r^{(t)})\le c(A_r^{(t-1)})+c(h)\le c(A_r^{(t-1)})+p$, while every other agent's cost weakly decreases. Since the sum of the costs increases by exactly $p$, all of these inequalities must hold with equality. Note that this argument allows $h$ to be an old chore.
\end{proof}
In other words, in this setting the agents' costs evolve exactly as the loads of identical machines in online makespan scheduling, and each agent's MMS equals the optimal makespan. Lower-bound constructions for this problem that present jobs in nondecreasing order of size therefore apply even under the bundle-level rule. Appendix~\ref{app:bundle-lower-bounds} proves lower bounds of $3/2$ for $n=2$, $5/3$ for $n=3$, and $\sqrt3$ for every $n\ge4$ (Theorem~\ref{thm:bundle-mms-lower-bound}), so round-robin insertion remains optimal for two and three agents.

Under this rule, the lower bound of $2-1/n$ still holds for ordinal algorithms, whose allocations depend only on the arrival history and the agents' rankings rather than numerical costs, and for algorithms maintaining EF1 (Theorems~\ref{thm:ordinal-chore-mms} and~\ref{thm:chore-ef1-mms-tradeoff}), even for identical strictly positive costs. Hence an improvement under bundle-level upgrades would have to use numerical costs and relax exact EF1; Theorem~\ref{thm:chore-mms-lower-bound} shows that even allowing both does not help under the item-level rule.

\section{Horizon Predictions}
\label{sec:predictions}
We now show that a small amount of future information improves on the optimal prediction-free factor in Theorem~\ref{thm:chore-mms-lower-bound}, without relaxing upgrade recourse. We also determine the best tradeoff between correct and arbitrary predictions for a natural class of periodic picking sequences.

\subsection{Reversing an Allocation of Cost Ranks}
List agent $i$'s costs for the chores in $D$ as $y_1^i\ge\cdots\ge y_m^i$, so that her cost at rank~$k$ is $y_k^i$; different agents' ranks may correspond to different chores. We set $y_k^i=0$ for $k>m$.

\begin{lemma}[Reverse least-cost picking]
\label{lem:reverse-chore-picking}
Let $\pi_1,\dots,\pi_m$ assign cost ranks to agents. Least-cost picking on $D$ in the reverse order $\pi_m,\dots,\pi_1$ gives agent $i$ cost at most $\sum_{k:\pi_k=i}y_k^i$.
\end{lemma}
\begin{proof}
At the turn corresponding to rank $k$, exactly $k$ chores remain, and among any $k$ chores, some chore costs agent $i$ at most her $k$th largest cost $y_k^i$. Her least costly choice at that turn therefore costs her at most $y_k^i$, and summing over her turns gives the claim.
\end{proof}
This reduction to cost ranks is standard in ordinal chore allocation~\citep{aziz2024ordinalchores}. Online, insertion can implement the reversed order, but since the first turn of this order corresponds to rank~$m$, choosing it requires information about the horizon; no information about the costs is needed.

\subsection{Only the Horizon Modulo a Short Period}
\citet{aziz2024ordinalchores} used a periodic assignment of cost ranks to agents in which each agent appears once or twice per period. We use the same period length $L=2n-\lfloor n/2\rfloor=n+\lceil n/2\rceil$, but change the positions of the repeated turns. The reason is that an incorrect prediction rotates the assignment of cost ranks: adjacent turns of the same agent may then give her the two most costly chores. Separating those turns limits this additional cost.

\begin{figure}[t]
\centering
\begin{tikzpicture}[turn/.style={circle,draw,fill=white,inner sep=0pt,minimum size=12pt,font=\small}]
\foreach \x/\name/\agents/\pairs/\vals in {0/2/{1,2,3,4,3,4}/{3/5,4/6}/{$C(2)=3/2$\\ $R(2)=7/3$},4.2/3/{1,2,3,4,2,3}/{2/5,3/6}/{$C(3)=5/3$\\ $R(3)=2$}}{
 \begin{scope}[xshift=\x cm]
  \draw[gray!60] (0,0) circle (0.7);
  \foreach \a [count=\k] in \agents {
   \node[turn] (P\k) at ({90-60*(\k-1)}:0.7) {$\a$};
   \node[gray,font=\scriptsize] at ({90-60*(\k-1)}:1.03) {\k};
  }
  \foreach \u/\v in \pairs {\draw[thick] (P\u) -- (P\v);}
  \node[font=\small,anchor=west,align=left] at (1.12,0) {$p^{(\name)}$:\\ \vals};
 \end{scope}
}
\end{tikzpicture}
\caption{Patterns $p^{(2)}$ and $p^{(3)}$ for $n=4$, with consistency $C$ and robustness $R$ from Theorem~\ref{thm:chore-residue-prediction}; positions are numbered clockwise, and lines join the two turns of each repeated agent. A correct prediction gives rank $k$ to position $k$; a wrong one starts elsewhere. In $p^{(2)}$, agent~$3$ has ranks $3$ and $5$, but a start at position~$3$ gives her ranks $1$ and $3$. In $p^{(3)}$, no start gives an agent two of the three most costly ranks, but agent~$2$ has ranks $2$ and $5$.}
\label{fig:patterns}
\end{figure}
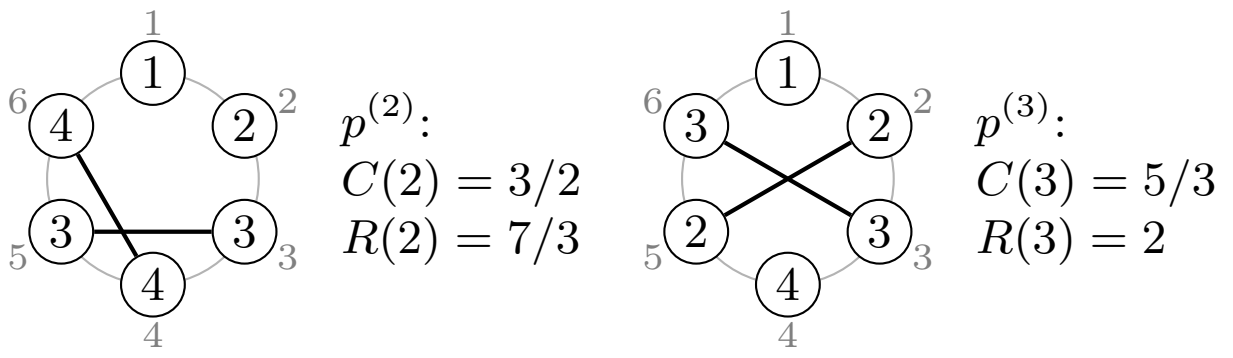

Choose an integer $\Delta$ with $\lfloor n/4\rfloor+1\le \Delta\le\lfloor L/2\rfloor$. The pattern $p^{(\Delta)}$ starts with $(1,\dots,n)$. If $\Delta\le\lceil n/2\rceil$, append $(n-\Delta+1,\dots,n)$ and then $(\lfloor n/2\rfloor+1,\dots,n-\Delta)$; otherwise, append $(n-\Delta+1,\dots,L-\Delta)$. Empty intervals are omitted. This gives $L$ turns, and the two occurrences of any repeated agent are at least $\Delta$ positions apart in either direction around the period. For example, when $n=4$, we have $p^{(2)}=(1,2,3,4,3,4)$ and $p^{(3)}=(1,2,3,4,2,3)$; see Figure~\ref{fig:patterns}.

Given a predicted residue $\widehat r\in\{0,\dots,L-1\}$ before the first arrival, run insertion with $\sigma_k=p^{(\Delta)}_{1+((\widehat r-k)\bmod L)}$ for $k \geq 1$, where remainders lie in $\{0,\dots,L-1\}$. With a correct residue, the ranks $1,2,\dots,m$ are thus assigned to the agents of $p^{(\Delta)}$ in order, cyclically. We call the worst MMS factor over all instances with $\widehat r=m\bmod L$ the \emph{consistency}, and the worst factor over all instances and predicted residues the \emph{robustness}. Consistency is thus guaranteed only at the final horizon and at earlier prefixes with the same residue, whereas robustness holds at every prefix, since every prefix is an instance with some residue.

\begin{theorem}
\label{thm:chore-residue-prediction}
For every integer $\Delta$ with $\lfloor n/4\rfloor+1\le \Delta\le\lfloor L/2\rfloor$, insertion with the sequence $\sigma$ above has upgrade recourse, and its consistency and robustness are exactly
\[
 C(\Delta)=\max\left\{1+\frac{n-1}{L},\frac32+\frac{\Delta-2}{L}\right\}
 \quad\text{and}\quad
 R(\Delta)=2+\frac{2(n-\Delta-1)}{L}.
\]
Both factors are attained on identical strictly positive costs.
\end{theorem}
\begin{proof}[Proof sketch]
Insertion ensures recourse for every prediction. An incorrect residue assigns the ranks according to a rotation of $p^{(\Delta)}$; by Lemma~\ref{lem:reverse-chore-picking}, each agent incurs at most the sum of her costs at her ranks, with equality for identical costs. Scale agent $i$'s costs so that $\MMS_i(D)=1$; by~\eqref{eq:mms-basic}, they sum to at most $n$ and each is at most $1$, and moreover $y_{n+1}^i\le1/2$. For consistency, an agent with one turn per period receives at most $1$ at her first rank and a $1/L$ fraction of the later costs. The worst repeated agent is $n-\Delta+1$, with ranks $n-\Delta+1$ and $n+1$: she may receive $1$ and $1/2$ there and a $2/L$ fraction of the later costs, which then total at most $(\Delta-2)/2$ for $\Delta\ge2$. For robustness, the two turns of a repeated agent are at least $\Delta$ positions apart around the period, so in the worst case she receives ranks $1$ and $\Delta+1$ and a $2/L$ fraction of the remaining costs. Appendix~\ref{app:predictions} gives the full proof and instances attaining both factors.
\end{proof}
Taking $\Delta=\lfloor n/4\rfloor+1$ yields consistency $1+(n-1)/L$ and robustness $2+2(n-\lfloor n/4\rfloor-2)/L<3$. The consistency is $4/3$ for $n=2$, $7/5$ for $n=3$, and less than $5/3$ for $n\ge4$, strictly improving on the optimal prediction-free factor in Theorem~\ref{thm:chore-mms-lower-bound}. Note that our insertion algorithm uses only pairwise comparisons and requires only ordinal information. Since $4/3$ for $n=2$ and $7/5$ for $n=3$ are optimal for deterministic ordinal algorithms even offline~\citep{liu1996ordinal,aziz2024ordinalchores}, our algorithm is optimal among ordinal ones for these values.

\subsection{The Optimal Consistency-Robustness Tradeoff}
Increasing $\Delta$ improves robustness but may worsen consistency (Figure~\ref{fig:patterns}). Our next result shows that no other pattern of the same length gives a better tradeoff.
\begin{theorem}
\label{thm:periodic-tradeoff}
Consider algorithms that fix a pattern $p\in[n]^L$ and use insertion with $\sigma_k=p_{1+((\widehat r-k)\bmod L)}$. No such algorithm has consistency below $1+(n-1)/L$. For every $\Gamma\ge1+(n-1)/L$, the minimum robustness among such algorithms with consistency at most $\Gamma$ is attained by $p^{(\Delta_\Gamma)}$ and equals
\begin{equation}
\label{eq:periodic-tradeoff}
 2+\frac{2(n-\Delta_\Gamma-1)}L,
 \quad\text{where }\Delta_\Gamma=\min\left\{\left\lfloor\frac L2\right\rfloor,
 \left\lfloor2+L\left(\Gamma-\frac32\right)\right\rfloor\right\}.
\end{equation}
\end{theorem}
\begin{proof}[Proof sketch]
One unit-cost chore followed by small chores of total cost $n-1$ gives the owner of rank~$1$ cost at least $1+(n-1)/L$. For $\Gamma\le C(\lfloor L/2\rfloor)<\min\{2,3n/L\}$, consistency forces distinct owners for the first $n$ ranks and at most two turns per agent per period, and an instance as in the proof of Theorem~\ref{thm:chore-residue-prediction} shows that the owner of rank $n+1$ has her first turn at most $\lfloor2+L(\Gamma-3/2)\rfloor$ ranks earlier. Her two turns are thus at most $\Delta_\Gamma$ positions apart around the period, and a suitable rotation forces robustness at least~\eqref{eq:periodic-tradeoff}. Larger values of $\Gamma$ are handled similarly, and Theorem~\ref{thm:chore-residue-prediction} gives the matching upper bound; see Appendix~\ref{app:predictions}.
\end{proof}
For $n=4$, Theorem~\ref{thm:periodic-tradeoff} shows that both patterns in Figure~\ref{fig:patterns} are optimal within the class. Since $\Delta_\Gamma\le\lfloor L/2\rfloor\le n-1$, the robustness in~\eqref{eq:periodic-tradeoff} is always at least $2$, above the prediction-free factor $2-1/n$: because $L>n$, some agent has two turns per period at most $n-1$ positions apart, and a wrong prediction can give her two of the $n$ most costly ranks. Hence, under arbitrary predictions, every algorithm in this class has a worse guarantee than round-robin insertion. Theorem~\ref{thm:periodic-tradeoff} does not rule out improvements using a different period, a nonperiodic sequence, or decisions that adapt to the observed numerical costs.

\section{EFX Requires Horizon Information}
\label{sec:efx}
We next contrast the MMS guarantees above with EFX. While the horizon modulo a short period suffices to improve the MMS factor for arbitrarily large horizons, for EFX, even uncertainty between two consecutive horizons can make every finite factor impossible.

\subsection{No Finite EFX Approximation}
\begin{theorem}
\label{thm:chore-efx-impossibility}
For every $n\ge2$ and finite $\alpha\ge1$, no deterministic prediction-free algorithm with upgrade recourse guarantees $\alpha$-EFX. Identical strictly positive costs suffice, and the final instance can admit an allocation that is exactly EFX and exactly MMS.
\end{theorem}
\begin{proof}
Choose $0<\eps<\min\{1/\alpha,1/2\}$ and present $n$ chores of common cost $\eps$. An $\alpha$-EFX guarantee at this stopping time forces one chore per agent: otherwise some agent has at least two chores, so her cost remains positive after deleting any one of them, while another agent has an empty bundle.

Now present a chore $z$ of common cost $1$, and let $r$ and $h$ be as in Lemma~\ref{lem:upgrade-sums}. Every agent other than $r$ finishes with cost at most $\eps$, so $z$ belongs to $r$ and each other agent holds at most one small chore. Hence $r$ also holds a small chore $d$. Removing $d$ leaves cost at least $1>\alpha\eps\ge\alpha c_r(A_j)$ for every $j\ne r$.

Offline, assign $z$ alone to one agent, two small chores to another, and one small chore to each remaining agent. This is EFX and has maximum cost equal to MMS $=1$.
\end{proof}
This contrasts with the offline setting, where a $2$-EFX allocation always exists for positive costs~\citep{garg2026efxchores}; the difficulty here comes from the online updates, not from the final instance. Whereas for goods, bundle-level upgrades permit exact EFX for identical valuations~\citep{huang2026upgrade}, for chores the impossibility persists under these upgrades even with identical costs, as we show at the end of this section.

\subsection{EFX with Horizon Information}
The preceding impossibility leaves open whether horizon information can help EFX as it helps MMS. We next determine both the best factor with the exact horizon and the amount of horizon information needed for any finite factor. In this subsection, EFX is required only at a final horizon consistent with the supplied information, and its factor may depend on $n$ and $m$, but not on the costs.

\begin{theorem}
\label{thm:efx-known-horizon}
For every $n\ge2$ and $m\ge1$, when the exact horizon $m$ is known, the optimal EFX factor with upgrade recourse is $\max\{1,m-n\}$. The upper bound is achieved by insertion in polynomial time. Identical strictly positive costs suffice for the lower bound.
\end{theorem}
\begin{proof}
If $m\le n$, use insertion with distinct owners for the first $m$ turns. Otherwise, put $K=m-n+1$ and use insertion with a sequence whose first $m$ picking turns are $K$ turns of agent~$1$ followed by one turn each of agents $2,\dots,n$. Agent $1$ receives her $K$ least costly chores and every other agent a singleton. After deleting any chore from agent $1$'s bundle, each of her $K-1$ remaining chores costs her no more than any other agent's singleton. This gives $(K-1)$-EFX; all other comparisons are immediate. Extend either sequence arbitrarily beyond $m$ to preserve recourse under an incorrect horizon prediction.

For the lower bound, suppose an algorithm guarantees some finite factor $\alpha$ on every length-$m$ instance, where $m>n$. Present $K$ chores of common cost~$1$. We claim that the algorithm must give all of them to one agent, even though EFX is not required at this time.

To see this, complete the instance with $n-1$ chores of common cost $H>\alpha K$. Since there are only $n-1$ such chores, some agent receives none of them and has final cost at most $K$. An owner of a cost-$H$ chore cannot have any additional chore, since deleting that additional chore would leave cost at least $H>\alpha K$. Thus these $n-1$ chores form singleton bundles, each of final cost $H$. By Lemma~\ref{lem:nondecreasing-chore-update}, each cost-$H$ arrival raises exactly one agent's cost by $H$, so each agent's final cost equals her cost after the unit-cost prefix plus a multiple of $H$. Since her cost after the prefix is at most $K<H$, a final cost of $H$ requires a cost of $0$ after the prefix. All $K$ unit-cost chores must therefore have been assigned to the remaining agent.

Now complete the same prefix with $n-1$ further unit-cost chores. The cost of the agent holding all $K$ unit-cost chores cannot decrease. Since $K\ge2$, any empty final bundle would violate every finite EFX factor. Each new arrival increases only one agent's cost, so the final costs must be $(K,1,\dots,1)$ up to permutation. Their EFX factor is $K-1=m-n$. Both completions have the same announced horizon.
\end{proof}
Thus exact horizon information makes a finite EFX approximation possible, but its optimal factor grows with the number of chores. Moreover, even a little uncertainty about the horizon can remove this possibility.

\begin{corollary}
\label{cor:efx-horizon-advice}
Suppose that, before the first arrival, an algorithm receives a message depending only on the horizon. If two horizons $2\le u<v$ with $v>n$ yield the same message, no finite EFX factor can be guaranteed at both. Consequently, if $1\le m\le M$, the minimum number of distinct messages that permit a finite EFX factor at every possible final horizon is $\max\{1,M-n+1\}$. In particular, knowing the horizon modulo a fixed period does not suffice when the horizon is unbounded.
\end{corollary}
\begin{proof}
Suppose that the same message is given for horizons $u$ and $v$. Applied at horizon $v$, the proof of Theorem~\ref{thm:efx-known-horizon} shows that at time $K=v-n+1$, a single agent holds all $K$ unit-cost chores. On the instance with $u$ unit-cost chores, the algorithm receives the same message and sees the same first $\min\{u,K\}$ arrivals. If $2\le u\le K$, then since no agent's cost ever decreases (Lemma~\ref{lem:nondecreasing-chore-update}), a single agent also holds all $u$ chores at time $u$. If $K<u<v$, then fewer than $n-1$ arrivals follow time $K$, each raising the cost of a single agent, so some agent's bundle remains empty while another agent has at least $K\ge2$ chores. In either case, no finite EFX factor holds at $u$.

Thus horizons $n+1,\dots,M$ require distinct messages, none of which can also serve horizon $2$. This gives the lower bound when $M>n$; otherwise one message is necessary. For the upper bound, use one message for all $m\le n$, assigning distinct owners, and a separate message specifying each $m>n$, using Theorem~\ref{thm:efx-known-horizon}.
\end{proof}
For example, knowing that $m\in\{u,u+1\}$ for $u\ge n$ is insufficient for any finite EFX approximation, in contrast to the $O(\log n)$ bits that improve MMS for arbitrarily large horizons.

Even exact knowledge of the horizon cannot reconcile a finite EFX factor with a nontrivial MMS factor that is independent of $m$.
\begin{corollary}
\label{cor:efx-mms}
Fix $n\ge2$ and $m\ge n+1$. For any algorithm that is given the exact horizon and guarantees a finite EFX factor at every horizon, there is an instance with $m$ chores and identical strictly positive costs on which some agent's MMS ratio is at least $(m-n+1)/\lceil m/n\rceil$. Consequently, no MMS factor $\beta<n$ independent of $m$ is compatible with such an EFX guarantee.
\end{corollary}
\begin{proof}
On the instance with $m$ unit-cost chores in the proof of Theorem~\ref{thm:efx-known-horizon}, some agent has cost $m-n+1$, whereas every agent's MMS equals $\lceil m/n\rceil$. Taking $m=nk$ gives ratio $n-(n-1)/k$, which tends to $n$. Since every allocation is $n$-MMS by~\eqref{eq:mms-basic}, this limit is tight.
\end{proof}
These lower bounds also hold for randomized algorithms whose EFX guarantee holds with probability one on every fixed input: since the cost $H$ in the proof does not depend on the random choices, distributing the initial unit-cost chores among several agents with positive probability would violate the guarantee on that fixed completion. For Corollary~\ref{cor:efx-mms}, this concerns the maximum realized MMS ratio, not each agent's expected cost.

\paragraph{EFX with bundle-level upgrades.}
The proof of Theorem~\ref{thm:chore-efx-impossibility} uses only Lemma~\ref{lem:upgrade-sums}, and the lower bounds with horizon information use only Lemma~\ref{lem:nondecreasing-chore-update}, which already allows bundle-level upgrades; since insertion satisfies the stronger item-level rule, the EFX characterization is unchanged.

\section{Conclusion}
In this paper, we have initiated the study of online fair division of chores with upgrade recourse. Round-robin insertion maintains EF1 and attains the optimal prediction-free MMS factor $2-1/n$, and $O(\log n)$ bits of horizon information improve on this factor when they are correct. By contrast, a finite EFX factor essentially requires the exact horizon, and even then it rules out every MMS factor below $n$ that does not depend on $m$. Overall, our results identify the horizon as the key information for chores: a few bits about it improve the optimal MMS factor, while its exact value is what makes a finite EFX factor possible.

\section*{Declaration of Generative AI Use}

The authors used GPT-6 as a research assistant when exploring proof ideas and revising the exposition. They
checked the mathematical arguments and take full responsibility for the contents of the paper.

\bibliographystyle{plainnat}
\bibliography{arxiv}
\clearpage
\appendix

\section{Further Related Work}
\label{app:further-related-work}
We expand here on the related work discussed in the introduction.

\paragraph{Online fair division.}
Online fair division allocates items under uncertainty about future arrivals~\citep{AleksandrovEtAl2015,AleksandrovWalsh2020Survey}. When every assignment is irrevocable, share-based guarantees are difficult to obtain. For chores, the trivial $n$-MMS guarantee is optimal up to an arbitrarily small additive term~\citep{song2025onlinechores}. For goods with general additive valuations, no positive approximation to MMS or EF1 is possible~\citep{neoh2026closing}, and even when each agent's valuation is normalized, no positive MMS guarantee is possible for three agents~\citep{zhou2023multi}. Randomization against an oblivious adversary, which fixes the input in advance, does not ensure these guarantees with high probability: for every $\alpha>0$, every randomized algorithm has an input on which it returns an $\alpha$-MMS allocation with probability at most $5/6$, and for every $\alpha\in(0,1]$ the corresponding probability for $\alpha$-EF1 is at most $e^{-\Omega(n)}$~\citep{cohen2026oblivious}. Under budget constraints, where fairness is evaluated only against budget-feasible subsets of each recipient's bundle, no deterministic online algorithm guarantees any fixed approximation to feasible envy-freeness without further structure~\citep{cohen2026budget}. In the related model in which agents arrive instead of items, no positive MMS guarantee is possible even for two agents without suitable information about the valuations of future agents~\citep{kulkarni2025online}. Previous work avoids these impossibilities in several ways: by weakening the fairness requirement, for example to envy that grows sublinearly in the number of goods~\citep{benade2018envy,zeng2020fairness}; by providing information about future goods~\citep{neoh2026additional,neoh2026frequency}; or by delaying assignments using a reordering buffer~\citep{amanatidis2026Online}. Temporal fair division gives the algorithm complete knowledge of future items and evaluates the allocation after every round; with irrevocable assignments, an allocation of goods that is EF1 after every round still need not exist, and deciding whether one exists is NP-hard~\citep{elkind2025temporal}. \citet{choi2026temporal} identify tractable cases for mixed manna, in which an item may be a good for some agents and a chore for others, and show that deciding whether an exact temporal MMS allocation exists is NP-hard for both goods and chores, even for two agents with identical valuations and two rounds. A related line of work minimizes the envy that accumulates as a sequence of items is allocated~\citep{goldberg2026cumulative}, or, when the same items are matched to the agents in every round, maximizes the minimum utility after every round~\citep{lim2026repeated}. We instead still assign each chore on arrival, but allow earlier assignments to change as long as no agent's earlier commitments become more costly for her.

\paragraph{Reassignment in dynamic fair division.}
Dynamic fair division usually measures the cost of reassignment by the number of changes. \citet{friedman2015dynamic} consider agents who arrive over time to share divisible resources, and bound the number of disruptions to earlier allocations. For indivisible goods arriving online, \citet{HeProcacciaPsomasZeng2019} maintain EF1 after every arrival by reassigning goods. In online house allocation, each agent holds exactly one house, so accepting a newly arrived house may require reassigning earlier ones; envy-freeability with subsidies can always be maintained using bounded recourse, although reassignment chains of length linear in the number of agents are unavoidable in the worst case~\citep{teh2026house}. Such bounds limit how often earlier assignments change, but not how much a single change raises an agent's cost. Upgrade recourse does the reverse: it places no bound on the number of reassignments, but each earlier chore of an agent must be replaced by a distinct chore that she finds no more costly. Insertion reassigns at most $m(m-1)/2$ chores in total (Section~\ref{sec:insertion}); we leave bounds on the number of reassignments under upgrade recourse to future work.

\paragraph{Recourse in online algorithms.}
Beyond fair division, online algorithms with \emph{recourse} may revise a limited number of earlier decisions. This model goes back to the dynamic Steiner tree problem~\citep{imase1991dynamic}; for online Steiner tree and minimum spanning tree, a constant competitive ratio can be maintained while changing only a constant number of edges per arrival, at least on average~\citep{megow2016power,gu2013power}. For online matching and related assignment problems, recourse counts the number of reassignments~\citep{gupta2014maintaining,bernstein2019online,angelopoulos2020online}. In scheduling and bin packing, the \emph{migration factor} bounds the total size of the jobs or items moved after an arrival relative to the size of the arriving one~\citep{sanders2009online,epstein2009robust,galvez2020symmetry}. \citet{boyar2022relaxing} instead restrict the type of change: an algorithm may accept an item after its arrival, or later reject an accepted item, but each such decision is final. In these problems, a solution is evaluated by a single objective, such as its makespan or the size of a matching. In fair division, each reassignment moves cost between particular agents, so a limit on the number or total size of changes does not prevent an agent from receiving chores far more costly than those she accepted. Upgrade recourse therefore restricts the type of change, as \citet{boyar2022relaxing} do, but separately for each agent, in the spirit of the non-worsening requirement for online metric embeddings~\citep{Coester2026Online}.

\paragraph{Offline chore allocation and EFX.}
The maximin share was introduced by \citet{budish2011combinatorial} and adapted to chores by \citet{aziz2017chores}; for chores, EF1 implies $(2-1/n)$-MMS~\citep{sun2023fairness}. A $2$-EFX allocation of chores always exists for positive additive costs~\citep{garg2026efxchores}, and an EFX allocation also exists when each chore has a size and each agent has a limited budget~\citep{elkind2024chorebudget}. Several works study instances in which the number of items exceeds the number of agents by at most a constant \citep{lim2026cost,neoh2025efx,teh2026surplus}. In our online model, by contrast, uncertainty between the horizons $n$ and $n+1$ already rules out every finite EFX factor (Theorem~\ref{thm:chore-efx-impossibility}).

\paragraph{Ordinal allocation and scheduling.}
With identical costs, a fixed picking sequence assigns sorted job ranks to machines, as in ordinal scheduling, where only the order of the job sizes is known. For makespan minimization on $m$ identical machines, \citet*{liu1996ordinal} give an ordinal algorithm with factor $1+(m-1)/(m+\lceil m/2\rceil)$, which is optimal for $m=2$ and $m=3$; for identical costs, the consistency $1+(n-1)/L$ in Theorem~\ref{thm:chore-residue-prediction} equals this factor. For chores with heterogeneous costs, \citet*{aziz2024ordinalchores} obtain ordinal MMS factors $4/3$, $7/5$, and $5/3$ for $n=2$, $n=3$, and $n\ge4$, respectively, the first two of which are optimal, and \citet{feigehuang2023chores} study picking sequences for chores more generally. Our contribution is to implement such sequences online with only the horizon modulo $L$ and to determine the optimal consistency--robustness tradeoff for fixed patterns; our prediction-free lower bound (Theorem~\ref{thm:chore-mms-lower-bound}) applies to algorithms that use revealed numerical costs. Our bundle-level lower bounds adapt lower-bound constructions for online makespan scheduling~\citep{faigle1989partition,rudin2003online,braun2025four}, which apply here because, for identical costs that are nondecreasing in arrival order, the agents' costs evolve as machine loads (Lemma~\ref{lem:nondecreasing-chore-update}).

\paragraph{Learning-augmented allocation.}
Predictions have been used for online Nash social welfare and more general allocation objectives~\citep{banerjee2022online,cohen2023general}. In indivisible fair division, \citet{neoh2026additional} transfer offline share guarantees using value-frequency information; with the same information, a positive approximation of EF1 is possible if and only if $n\le4$, and EFX admits no positive approximation for three or more agents~\citep{neoh2026frequency}. Other work studies proportionality up to one good~\citep{choo2026proportionality,neoh2026closing}, EFX~\citep{melissourgos2025onlineefx}, envy-freeness under budget constraints~\citep{cohen2026budget}, and subsidy minimization in online house allocation~\citep{teh2026house}. These works concern goods and predict future values. For chores, our predictions concern only the number of chores modulo a short period, and we determine the optimal tradeoff between consistency and robustness among fixed patterns of length $L$.

\paragraph{Temporal voting and other sequential collective decisions.}
Fairness across a sequence of collective decisions is also the focus of temporal (or perpetual) voting, in which a fixed group of voters makes a decision in each of several rounds and rules may take earlier decisions into account~\citep{lackner2020perpetual}; \citet{elkind2024temporal} propose a unified framework for these models. Earlier dynamic models choose an alternative in each round under evolving preferences or to maximize long-term Nash welfare~\citep{parkes2013dynamic,freeman2017dynamic}. In public decision making, several independent issues are decided at once~\citep{conitzer2017fair,skowron2022proportional,masarik2024generalised}, a model that generalizes the allocation of indivisible goods~\citep{conitzer2017fair}. Much of the temporal literature concerns proportionality. Perpetual voting rules have been analyzed axiomatically~\citep{lackner2020perpetual,lackner2023proportional}, and justified representation and its variants have been adapted to the temporal setting~\citep{bulteau2021justified,chandak2024proportional} and strengthened beyond EJR, for example to EJR+ and FJR, which remain satisfiable~\citep{phillips2026strengthening}. Verifying such axioms is harder than in multiwinner voting~\citep{elkind2025verifying}, although a strengthening of FJR can be both computed and verified in polynomial time~\citep{teh2026fjr}. Other work considers ranked preferences~\citep{hazon2026ranked}, bounds how often a voter is dissatisfied~\citep{kozachinskiy2025optimal}, and adapts the maximin share to binary perpetual decisions~\citep{gafni2024beyond}. For welfare, maximizing utilitarian welfare is easy, whereas optimizing egalitarian welfare is intractable even in restricted cases~\citep{elkind2024welfare}, including over \emph{public chores}, which benefit society but impose costs on the individuals affected, where online algorithms have also been analyzed through their competitive ratio~\citep{elkind2025chores}. Enforcing proportionality can incur a welfare loss that grows sublinearly with the number of voters or rounds~\citep{teh2026price}. Related models assign projects to time slots~\citep{elkind2022slot} or elect sequences of committees that bound how often candidates serve~\citep{bredereck2020successive}, how much consecutive committees differ~\citep{bredereck2022votes}, or how often each agent is satisfied~\citep{deltl2023algorithmics}; \citet{zech2024aversion} seek a second-stage winning committee that overlaps as much as possible with the first. As in our setting, it matters how much of the future is known: among proportional rules for sequential decisions, Sequential Phragm\'en decides online, whereas the Method of Equal Shares and Proportional Approval Voting obtain stronger guarantees semi-online and offline, respectively~\citep{chandak2024proportional}; with ranked preferences, some guarantees require all future preferences, while others need only the number of rounds or no information about the future~\citep{hazon2026ranked}. Two features distinguish temporal voting from our model. Each decision there is a public outcome shared by all voters, whereas each chore here is performed by one agent. Moreover, earlier decisions are typically not revised, or changes between consecutive outcomes are measured by their size, whereas upgrade recourse permits any change that does not make an agent's earlier commitments more costly for her.

\section{Omitted Proofs in Section~\ref{sec:lower-bounds}}
\label{app:bundle-lower-bounds}
We prove the bundle-level MMS bounds summarized in Section~\ref{sec:bundle-mms}. All constructions use identical costs and Lemma~\ref{lem:nondecreasing-chore-update}.

\subsection{Proof of Theorem~\ref{thm:bundle-mms-lower-bound}}
We begin with lower bounds that impose no additional restriction on the algorithm.
\begin{theorem}
\label{thm:bundle-mms-lower-bound}
No deterministic prediction-free algorithm with bundle-level upgrade recourse guarantees an MMS factor below $3/2$ for $n=2$, $5/3$ for $n=3$, or $\sqrt3$ for any $n\ge4$. The bounds hold even for identical strictly positive costs.
\end{theorem}
\begin{proof}
For $n=2$, present costs $1,1,2$. A factor below $3/2$ forces cost $1$ on each agent after the first two arrivals, since otherwise one agent has cost $2$ while MMS is $1$. Lemma~\ref{lem:nondecreasing-chore-update} then gives cost $3$ to one agent, while final MMS is~$2$.

For $n=3$, present $1,1,1,3,3,3,6$. A factor below $5/3$ forces initial cost $1$ on each agent. After the three cost-$3$ arrivals, MMS is $4$, and it is at most $4$ throughout these arrivals; hence any agent receiving two such cost increases would have cost at least $7>(5/3)4$ at that time. Thus all three costs are $4$. The last arrival gives one agent cost $10$, whereas the partition $\{6\}$, $\{3,3\}$, $\{3,1,1,1\}$ has maximum cost $6$, equal to MMS. These small-agent constructions are due to \citet{faigle1989partition}.

For $n\ge4$, we extend the four-machine constructions of \citet{rudin2003online} and \citet{braun2025four}, as follows.

For each $\beta$ with $1+1/\sqrt2<\beta<\sqrt3$, we rule out every MMS factor strictly below $\beta$. This suffices because $\beta$ can be arbitrarily close to $\sqrt3$. By~\eqref{eq:nondecreasing-chore-update}, each arrival increases exactly one agent's cost, by the new chore's cost; we refer to these cost increases, not to the ownership of individual chores.

\paragraph{Construction.}
Let $q:=\lfloor n/4\rfloor\ge1$, so $0\le n-4q\le3$. Set $s_0:=(\beta-1)/(2-\beta)$ and $b_0:=1+s_0$, and note that $s_0>1+\sqrt2>2$ since $\beta>1+1/\sqrt2$. First present $n$ chores of cost $1$, followed by $n$ chores of cost $s_0$. For each $k\ge1$, define
\begin{equation}
\label{eq:chore-recursive-costs}
    s_k:=\frac{2b_{k-1}}{3\beta-5},\quad
    p_k:=\frac{s_k}{2}-2s_{k-1},\quad
    b_k:=b_{k-1}+p_k+s_k.
\end{equation}
Block $k$ consists, in order, of $n$ chores of cost $p_k$, $n-q$ chores of cost $s_k$, and $q$ chores of cost $s_k+2s_{k-1}$. Every agent's cost after this block will be at least $b_k$, because the algorithm must give the first $n$ and the last $n$ cost increases of the block to distinct agents. We will show that some finite $K$ satisfies $b_K\ge2(\beta-1)s_K$. After block $K$, present one last chore of cost $2s_K$. At that time, the earlier chores will fit into at most $n-1$ bundles of cost at most $2s_K$, so the last chore alone determines MMS. Arrivals may instead stop earlier as soon as an agent's cost is at least $\beta$ times her current MMS.

\paragraph{Bounds on the costs and the number of blocks.}
To compare the accumulated cost with the last chore, write $x_k:=b_k/s_k$. The definitions of $s_0,b_0$ and~\eqref{eq:chore-recursive-costs} give
\begin{equation}
\label{eq:chore-cost-ratio}
    x_0=\frac1{\beta-1},\quad
    x_k=\frac{3\beta-2}{2}-\frac{3\beta-5}{x_{k-1}}.
\end{equation}
We have $0<3\beta-5<1/5$ and $15/11<x_0<3/2$. Moreover, $x_1-x_0={(2-\beta)(3\beta-5)(2\beta-1)}/{(2(\beta-1))}>0$. The right-hand side of~\eqref{eq:chore-cost-ratio} is increasing in $x_{k-1}$ and, at $x_{k-1}=3/2$, equals $3/2-(3\beta-5)/6<3/2$. Induction therefore shows that $x_k$ is strictly increasing and
\begin{equation}
\label{eq:chore-cost-estimates}
    \frac{15}{11}<x_k<\frac32,\quad
    \frac{p_k}{s_{k-1}}=
    \frac{x_{k-1}}{3\beta-5}-2>\frac{53}{11}
    \quad \text{for } k\ge1.
\end{equation}
In particular, all costs are positive and nondecreasing in arrival order. Indeed, $2s_{k-1}<p_k<s_k<s_k+2s_{k-1}<2s_k$, so every chore of block $k$ costs at least $p_k$ and less than $2s_k$, while every earlier chore costs less than $2s_{k-1}$ (for $k=1$, the earlier costs are $1$ and $s_0$). Thus~\eqref{eq:nondecreasing-chore-update} applies to every arrival, including the last chore.

It remains to justify a finite $K$. Initially $x_0<2(\beta-1)$. If all $x_k$ stayed below $2(\beta-1)$, their increasing sequence would converge to a fixed point of~\eqref{eq:chore-cost-ratio} in $[x_0,2(\beta-1)]$. But the quadratic $-x^2+(3\beta-2)x/2-(3\beta-5)$, which vanishes exactly at these fixed points, is concave and positive at both endpoints: at $x_0$ its value is $x_0(x_1-x_0)>0$, and at $2(\beta-1)$ it is $3-\beta^2>0$. It is therefore positive throughout the interval, so no such fixed point exists. Hence $x_K\ge2(\beta-1)$ for some finite $K$.

\paragraph{Partitions of the chores.}
The following grouping is used only to construct partitions for bounding MMS. Divide the chores into $q$ groups of the first type and $n-4q$ groups of the second type. A group of the first type contains four unit-cost chores and four cost-$s_0$ chores initially, and in each block $k$ it contains four cost-$p_k$ chores, three cost-$s_k$ chores, and one chore of cost $s_k+2s_{k-1}$. A group of the second type contains one unit-cost chore and one cost-$s_0$ chore initially, and one chore of each cost $p_k,s_k$ per block. These groups account for every chore exactly once. A group of the second type has total cost $b_k$ after block $k$.

Each group of the first type can be partitioned into three bundles of cost at most $2s_k$ after block $k$. Initially, use two pairs of cost-$s_0$ chores and a bundle of four unit-cost chores; all fit since $s_0>2$. For the induction, suppose its earlier chores have been partitioned into $P_1,P_2,P_3$, each of cost at most $2s_{k-1}$. Form three new bundles as follows: the first contains two cost-$s_k$ chores; the second contains $P_1\cup P_2$, one cost-$s_k$ chore, and two cost-$p_k$ chores; the third contains $P_3$, the chore of cost $s_k+2s_{k-1}$, and two cost-$p_k$ chores. The three costs are at most $2s_k$ because $2p_k+4s_{k-1}=s_k$ by~\eqref{eq:chore-recursive-costs}.

We next bound $\MMS_i(D^{(t)})$ for every agent $i$ at the intermediate stopping times needed below.

After a complete block $k$, there is an $n$-partition with every bundle of cost at most $3s_k/2$. For each group of the first type, take its three earlier bundles $P_1,P_2,P_3$ and add one chore of each cost $p_k,s_k$ to each. Put its remaining two chores, of costs $p_k$ and $s_k+2s_{k-1}$, in a fourth bundle. All four bundles cost at most $3s_k/2$ because $p_k+s_k+2s_{k-1}=3s_k/2$ by~\eqref{eq:chore-recursive-costs}. Each group of the second type fits in one bundle since $b_k<3s_k/2$. The initial chores also admit an $n$-partition of cost $b_0<3s_0/2$ per bundle, by pairing each unit-cost chore with a cost-$s_0$ chore.

After the first $n$ arrivals of block $k$, add one cost-$p_k$ chore to each bundle of the preceding $n$-partition. This gives $\MMS_i(D^{(t)})\le p_k+3s_{k-1}/2$. After the next $n-q$ arrivals, form $n-q$ pairs of costs $p_k+s_k$. Put all earlier chores from each group of the first type into one of $q$ further bundles, and put all the earlier chores in groups of the second type into the first of these bundles. There are at most three such groups. The earlier total cost in any of these $q$ bundles is therefore at most $6s_{k-1}+3b_{k-1}<9b_{k-1}<s_k$, where we used~\eqref{eq:chore-cost-estimates} and $3\beta-5<1/5$. Adding one of the $q$ remaining cost-$p_k$ chores to each gives an $n$-partition showing that $\MMS_i(D^{(t)})\le p_k+s_k$. Each of these two bounds also applies at earlier arrivals within its corresponding part of the block: delete the not-yet-arrived chores from the partition. During the last $q$ arrivals, the complete-block bound $3s_k/2$ applies in the same way.

Finally, the three-bundle partitions of the $q$ groups of the first type, together with one bundle per group of the second type, place all chores through block $K$ in $3q+(n-4q)=n-q\le n-1$ bundles of cost at most $2s_K$. Put the final chore alone, adding empty bundles if needed. Its cost also lower-bounds MMS, so $\MMS_i(D)=2s_K$ for every agent~$i$.

\paragraph{Costs required of the online algorithm.}
Suppose the algorithm guarantees an MMS factor strictly below $\beta$. The first $n$ unit-cost increases must go to distinct agents, since otherwise some agent has cost at least $2$ when MMS is at most $1$. The next $n$ cost-$s_0$ increases must likewise go to distinct agents: at every such arrival MMS is at most $1+s_0$, and $(1+2s_0)/(1+s_0)=\beta$. Thus every agent has cost $b_0$ after these initial arrivals.

Assume every agent has cost at least $b_{k-1}$ before block $k$. If two cost-$p_k$ increases go to the same agent, then at the second of them, the bound $\MMS_i(D^{(t)})\le p_k+3s_{k-1}/2$ gives her a ratio of at least
\[
    \frac{b_{k-1}+2p_k}{p_k+\frac32s_{k-1}}
    >\frac{15/11+2(53/11)}{53/11+3/2}
    =\frac{242}{139}>\sqrt3>\beta.
\]
The strict inequality follows by substituting the lower bounds from~\eqref{eq:chore-cost-estimates}, since the ratio increases with $b_{k-1}$, and also with $p_k$ because $b_{k-1}<3s_{k-1}$. Hence each agent receives exactly one cost-$p_k$ increase.

If an agent receives two of the next $n-q$ cost increases, then, by the bound $\MMS_i(D^{(t)})\le p_k+s_k$, her ratio is at least $(b_{k-1}+p_k+2s_k)/(p_k+s_k)>\beta$. Adding $2s_{k-1}$ to its numerator and denominator strictly decreases this ratio, since its numerator exceeds its denominator, and gives
\begin{equation}
\label{eq:chore-repeated-increase}
    \frac{b_{k-1}+p_k+2s_k+2s_{k-1}}
         {p_k+s_k+2s_{k-1}}
    =\frac{b_{k-1}+(5/2)s_k}{(3/2)s_k}
    =\beta.
\end{equation}
Here the first equality uses $p_k+2s_{k-1}=s_k/2$ and the second uses $s_k=2b_{k-1}/(3\beta-5)$, both from~\eqref{eq:chore-recursive-costs}. If one of the last $q$ increases goes to an agent who has already received a cost-$s_k$ or cost-$(s_k+2s_{k-1})$ increase in this block, her ratio is at least the left-hand side of~\eqref{eq:chore-repeated-increase}, using the complete-block MMS bound, which again contradicts the assumed guarantee. Consequently, the last $n$ increases of the block go to distinct agents, and every agent ends with cost at least $b_{k-1}+p_k+s_k=b_k$.

Hence, after block $K$, every agent has cost at least $b_K$. The final arrival increases one agent's cost by $2s_K$. Since $\MMS_i(D)=2s_K$, her ratio is at least $1+b_K/(2s_K)\ge1+(\beta-1)=\beta$, a contradiction. Every earlier contradiction also occurs at a valid finite stopping time.
\end{proof}

\subsection{Proof of Theorem~\ref{thm:ordinal-chore-mms}}
Next, we consider ordinal algorithms.

\begin{theorem}
\label{thm:ordinal-chore-mms}
For every $n\ge2$, the optimal MMS factor of deterministic ordinal algorithms with bundle-level upgrade recourse and without predictions is $2-1/n$, even without EF1. Identical strictly positive costs suffice for the lower bound.
\end{theorem}
\begin{proof}
Use common arrival costs $1,2,4,\dots$. Distinct subsets have distinct costs, so the equalities in~\eqref{eq:nondecreasing-chore-update} force each update to add the arriving chore to one bundle and leave every old chore with its owner. An ordinal algorithm must make the same assignments on \emph{every} common cost sequence strictly increasing in arrival order, as in the corresponding argument for goods~\citep{huang2026upgrade}.

For an integer $k\ge2$, let $m$ be the first time some agent receives her $k$th chore; then $m-1\le n(k-1)$. For $\eps>0$, change the first $m-1$ costs to strictly increasing values in $[1,1+\eps)$ and set the last cost to $p=(1+\eps)\lceil(m-1)/(n-1)\rceil$. Rankings and assignments are unchanged. The last recipient incurs at least $p+k-1$. MMS equals $p$: put the last chore alone and distribute the others among $n-1$ bundles with at most $\lceil(m-1)/(n-1)\rceil$ chores each. Her ratio is therefore at least $1+(k-1)/p$. Since $m-1\le n(k-1)$ and $\lceil x\rceil\le x+1$, we have $p\le(1+\eps)(nk-1)/(n-1)$, so letting $k\to\infty$ and then $\eps\to0$ gives the lower bound $1+(n-1)/n=2-1/n$. Round-robin insertion is ordinal and matches this lower bound.
\end{proof}

\subsection{Proof of Theorem~\ref{thm:chore-ef1-mms-tradeoff}}
The next bound, relating approximate EF1 to MMS, is the chore counterpart of an item-level bound for goods~\citep{huang2026upgrade}, but it holds under the weaker bundle-level rule.
\begin{theorem}
\label{thm:chore-ef1-mms-tradeoff}
If a deterministic prediction-free algorithm with bundle-level upgrade recourse guarantees both $\alpha$-EF1 and $\beta$-MMS, where $\alpha\ge1$, then $\beta\ge1+(n-1)/(1+\alpha(n-1))$. In particular, with exact EF1, the optimal MMS factor is $2-1/n$.
\end{theorem}
\begin{proof}
Present $M$ common unit-cost chores, where $M$ is a multiple of $n-1$. Let $a$ be the minimum bundle size after these arrivals. Every other bundle has at most $\alpha a+1$ chores by $\alpha$-EF1, so $a\ge(M-(n-1))/(1+\alpha(n-1))$. Append a chore of cost $M/(n-1)$. Lemma~\ref{lem:nondecreasing-chore-update} forces final cost at least $a+M/(n-1)$ on some agent. MMS is $M/(n-1)$, attained by placing the last chore alone and evenly distributing the unit chores among $n-1$ parts. Her ratio is therefore at least $1+(n-1)a/M$. By the bound on $a$, the limit inferior of $(n-1)a/M$ as $M\to\infty$ is at least $(n-1)/(1+\alpha(n-1))$, proving the claimed bound on~$\beta$. For $\alpha=1$, round-robin insertion achieves equality.
\end{proof}

\section{Omitted Proofs in Section~\ref{sec:predictions}}
\label{app:predictions}

\subsection{Proof of Theorem~\ref{thm:chore-residue-prediction}}
\begin{proof}
Insertion ensures recourse for every prediction, and an incorrect residue assigns the ranks according to a rotation of $p^{(\Delta)}$. By Lemma~\ref{lem:reverse-chore-picking}, each agent's cost is at most her \emph{rank sum} (the sum of her costs at the ranks assigned to her) with equality for identical costs.

\paragraph{Consistency.}
Fix an agent $i$ and scale her costs so that $\MMS_i(D)=1$; if her MMS is zero, the claim is immediate. Then~\eqref{eq:mms-basic} and an MMS partition give
\begin{equation}
\label{eq:periodic-rank-bounds}
 \sum_{k\ge1}y_k^i\le n,\quad y_1^i\le1,\quad y_{n+1}^i\le\frac12.
\end{equation}
For the last inequality, $n+1$ chores each costing more than $1/2$ cannot fit into $n$ bundles of cost at most $1$.

We use the averaging inequality of \citet{aziz2024ordinalchores}: if $z_1\ge\cdots\ge z_L\ge0$ and $L/2\le b\le L$, then $z_b+z_L\le(2/L)\sum_{k=1}^Lz_k$. Indeed, the sum is at least $bz_b+(L-b)z_L\ge L(z_b+z_L)/2$.

An agent appearing once per period first appears at rank $i$. Her rank sum is at most
\begin{align*}
 y_i^i+\frac1L\sum_{k>i}y_k^i
 &\le\frac nL+\left(1-\frac iL\right)y_i^i \le1+\frac{n-i}{L}\le1+\frac{n-1}{L}.
\end{align*}
Here each later assigned rank is the last in a block of $L$ ranks, so it costs at most the average of that block; we also used $\sum_{k>i}y_k^i\le n-iy_i^i$ and $y_i^i\le1$, both from~\eqref{eq:periodic-rank-bounds}.

Next, suppose that $i$ belongs to the final appended interval. Then $\lfloor n/2\rfloor+1\le i\le n-\Delta$ and $\Delta<\lceil n/2\rceil$. Her two positions differ by $\lceil n/2\rceil+\Delta>L/2$. Applying the averaging inequality to each block of $L$ ranks after rank $i$ gives rank sum at most
\begin{align*}
 y_i^i+\frac2L\sum_{k>i}y_k^i
 &\le\frac{2n}{L}+\left(1-\frac{2i}{L}\right)y_i^i\\
 &\le\max\left\{\frac{2n}{L},1+\frac{2(n-i)}L\right\}
 \le1+\frac{n-1}{L},
\end{align*}
where the second inequality follows by considering the sign of $1-2i/L$, and the last uses $L\ge n+1$, $i\ge\lfloor n/2\rfloor+1$, and $2(\lceil n/2\rceil-1)\le n-1$.

Every other repeated agent has first two ranks $i$ and $j=i+\Delta$, where $i\ge n-\Delta+1$ and $j\ge n+1$. In each block of $L$ ranks after $j$, her assigned positions are $L-\Delta$ and $L$. Since $\Delta\le L/2$, averaging those blocks and using $\sum_{k\le j}y_k^i\ge i y_i^i+\Delta y_j^i$ gives rank sum at most $2n/L+(1-2i/L)y_i^i+(1-2\Delta/L)y_j^i$. If $1-2i/L\ge0$, use $y_i^i\le1$ and $y_j^i\le1/2$; the bound becomes $3/2+(2n-2i-\Delta)/L\le3/2+(\Delta-2)/L$. If $1-2i/L<0$, use $y_i^i\ge y_j^i$ instead, so that the combined coefficient of $y_j^i$ becomes $2(1-j/L)\ge0$; with $y_j^i\le1/2$, the bound is then at most $1+(2n-j)/L\le1+(n-1)/L$. This proves the consistency upper bound $C(\Delta)$.

\paragraph{Robustness.}
The two occurrences of every repeated agent are at least $\Delta$ positions apart in either direction around the period. For agents in the first appended interval, the distances are $\Delta$ and $L-\Delta$. For agents in the final appended interval, they are $\lceil n/2\rceil+\Delta$ and $n-\Delta$; that interval is nonempty only when $\Delta<\lceil n/2\rceil$, so $n-\Delta\ge \Delta$. Agent $n-\Delta+1$ has distance exactly $\Delta$.

For any rotation, let $N_i(k)$ count agent $i$'s occurrences among the first $k$ ranks. Then
\begin{equation}
\label{eq:periodic-counts}
 N_i(k)\le
 \begin{cases}
 1,&1\le k\le \Delta,\\
 2+\dfrac2L(k-\Delta-1),&k\ge \Delta+1.
 \end{cases}
\end{equation}
The first inequality holds because any $\Delta$ consecutive ranks contain at most one occurrence of each agent. To verify the second inequality, write $k=aL+b$ with $0\le b<L$. The complete periods contribute at most $2a$ occurrences. The remaining positions contribute at most $0$, $1$, or $2$, according as $b=0$, $1\le b\le \Delta$, or $b\ge \Delta+1$. Substitution, using $\Delta\le L/2$, gives~\eqref{eq:periodic-counts} in each case.

Agent $i$'s rank sum equals $\sum_{k\ge1}N_i(k)(y_k^i-y_{k+1}^i)$ by summation by parts. Since each difference $y_k^i-y_{k+1}^i$ is nonnegative, replacing $N_i(k)$ by the right-hand side of~\eqref{eq:periodic-counts} and summing by parts again shows that the rank sum is at most
\begin{align*}
 y_1^i+y_{\Delta+1}^i+\frac2L\sum_{k>\Delta+1}y_k^i&\le\frac{2n}{L}+\left(1-\frac2L\right)y_1^i+\left(1-\frac{2\Delta}{L}\right)y_{\Delta+1}^i\\
 &\le2+\frac{2(n-\Delta-1)}L.
\end{align*}
For the first inequality, we bound the sum over $k>\Delta+1$ using $\sum_{k\ge1}y_k^i\le n$ from~\eqref{eq:periodic-rank-bounds} together with $\sum_{k=1}^{\Delta+1}y_k^i\ge y_1^i+\Delta y_{\Delta+1}^i$. The second holds because both coefficients are nonnegative, $y_{\Delta+1}^i\le y_1^i$, and $y_1^i\le1$ by~\eqref{eq:periodic-rank-bounds}.

\paragraph{Matching instances.}
All costs in the following constructions are identical and strictly positive. To attain $1+(n-1)/L$, take one unit-cost chore and $M$ equal chores of cost $(n-1)/M$, where $M$ is a sufficiently large common multiple of $L$ and $n-1$. MMS is $1$: put the unit chore alone and split the rest equally among $n-1$ bundles. Agent $1$ appears once per period and receives cost $1+(n-1)/L$ under the correct residue.

For $\Delta\ge2$, let $i=n-\Delta+1$. Take $i$ unit-cost chores, $\Delta$ half-unit-cost chores, and equal small chores of total cost $(\Delta-2)/2$. When $\Delta>2$, choose their number to be a sufficiently large common multiple of $L$ and $\Delta-2$; for $\Delta=2$, omit them. MMS is $1$: use $i$ singleton unit bundles, pair the half-unit chores where possible, and fill the remaining capacity in the other $\Delta-1$ bundles with small chores. Their common cost divides $1/2$, so this is possible. Agent $i$ receives a unit at rank $i$, a half-unit at rank $n+1$, and a $2/L$ fraction of the small chores. Her cost is $3/2+(\Delta-2)/L$. For $\Delta=1$, the first term of $C(\Delta)$ is larger, so the first construction suffices.

To attain $R(\Delta)$, take $\Delta+1$ unit-cost chores and equal small chores of total cost $n-\Delta-1$. This total is nonnegative because $\lfloor L/2\rfloor\le n-1$. If it is positive, choose the number of small chores to be a sufficiently large common multiple of $L$ and $n-\Delta-1$; otherwise omit them. MMS is $1$. Choose a residue that rotates agent $n-\Delta+1$'s two occurrences to ranks $1$ and $\Delta+1$. She receives the unit-cost chores at ranks $1$ and $\Delta+1$ and a $2/L$ fraction of the small chores, giving cost $2+2(n-\Delta-1)/L$. Thus both stated factors are exact.
\end{proof}

\subsection{Proof of Theorem~\ref{thm:periodic-tradeoff}}
\begin{proof}
For any length-$L$ pattern, use one unit-cost chore followed by equal small chores of total cost $n-1$, with their number a sufficiently large common multiple of $L$ and $n-1$. MMS is $1$. The owner of rank $1$ appears at least once per period and receives cost at least $1+(n-1)/L$. This proves the minimum consistency claim.

First suppose that $1+(n-1)/L\le\Gamma\le C(\lfloor L/2\rfloor)$. We have $C(\lfloor L/2\rfloor)\le\max\{1+(n-1)/L,\,2-2/L\}<\min\{2,3n/L\}$, where the last inequality uses $n<L\le(3n+1)/2$. Thus the first $n$ ranks of any $\Gamma$-consistent pattern must have distinct owners: otherwise, $n$ unit-cost chores give an agent cost $2$ while MMS is $1$. Relabel the agents so that these ranks have owners $1,\dots,n$. No agent can appear three times per period, since on $kL$ unit-cost chores her MMS ratio would tend to at least $3n/L>\Gamma$ as $k\to\infty$. Hence every agent appears once or twice.

Suppose the owner of rank $n+1$ first appears at rank $i$, and put $g=n+1-i$. For $g\ge2$, take $i$ unit-cost chores, $g$ half-unit-cost chores, and small chores totaling $(g-2)/2$, as in the matching consistency construction above. MMS is $1$, and this agent receives cost $3/2+(g-2)/L$. Consistency therefore requires $g\le\lfloor2+L(\Gamma-3/2)\rfloor$. The same inequality holds for $g=1$: since $\Gamma\ge1+(n-1)/L$, we have $2+L(\Gamma-3/2)\ge n+1-L/2=1+\lfloor n/2\rfloor/2$, so its right-hand side is at least $\lfloor n/4\rfloor+1$.
The smaller distance $\delta$ between these two appearances around the period is at most both $g$ and $\lfloor L/2\rfloor$, so $\delta\le \Delta_\Gamma$. Rotate this pair to ranks $1$ and $\delta+1$. The construction with $\delta+1$ unit-cost chores and equal small chores of total cost $n-\delta-1$ then forces robustness at least $2+2(n-\delta-1)/L\ge2+2(n-\Delta_\Gamma-1)/L$.

Now suppose that $\Gamma>C(\lfloor L/2\rfloor)$. Then $\Gamma>3/2+(\lfloor L/2\rfloor-2)/L$, so $\Delta_\Gamma=\lfloor L/2\rfloor$, and we show that no length-$L$ pattern has robustness below $R(\lfloor L/2\rfloor)$, even without a consistency requirement. If each agent appears at most twice, some agent appears twice because $L>n$. The smaller distance between her appearances is at most $\lfloor L/2\rfloor$, and the preceding construction applies. If an agent appears at least three times, rotate an occurrence to rank $1$ and use the construction with one unit-cost chore. She receives cost at least $1+3(n-1)/L$, and $1+3(n-1)/L-R(\lfloor L/2\rfloor)=(n+2\lfloor L/2\rfloor-L-1)/L\ge0$ because $2\lfloor L/2\rfloor\ge L-1$ and $n\ge2$.

Finally, let $\Gamma\ge1+(n-1)/L$. As shown above, $\lfloor2+L(\Gamma-3/2)\rfloor\ge\lfloor n/4\rfloor+1$, so $\Delta_\Gamma$ lies in the allowed range, and $\Delta_\Gamma\le2+L(\Gamma-3/2)$ gives $C(\Delta_\Gamma)\le\Gamma$. Theorem~\ref{thm:chore-residue-prediction} therefore gives the matching upper bound using $p^{(\Delta_\Gamma)}$.
\end{proof}

\end{document}